\documentclass[a4paper]{jpconf}

\usepackage{graphicx}
\usepackage{citesort}
\begin{document}
\nocite{*}
\title{Search for rare $b$-hadron decays at CDF}

\author{J. Heuser and M. Milnik for the CDF Collaboration}

\address{Institut f\"ur Experimentelle Kernphysik, University of Karlsruhe, Wolfgang-Gaede-Str. 1, 76131 Karlsruhe, Germany}

\ead{heuser@ekp.uka.de, milnik@ekp.uka.de}

\begin{abstract}
We report on searches for
$B^0_s\rightarrow\mu^+\mu^-$, $B^0_d\rightarrow\mu^+\mu^-$ decays
and
$b\rightarrow s \mu^+\mu^-$ transitions
in exclusive decays of B mesons.
Using 2 fb$^{-1}$ of data collected by the CDF II detector we find upper limits on the
branching fractions
$\mathcal{B}(B^0_s\rightarrow\mu^+\mu^-)<5.8\times 10^{-8}$ and
$\mathcal{B}(B^0_d\rightarrow\mu^+\mu^-)<1.8\times 10^{-8}$ at 95$\%$ confidence level.
Using 924 pb$^{-1}$ of data we measure the branching fractions
$\mathcal{B}(B^+\rightarrow \mu^+\mu^-K^+)=(0.60\pm0.15\pm0.04)\times 10^{-6}$,
$\mathcal{B}(B^0_d\rightarrow \mu^+\mu^-K^{*0})=(0.82\pm0.31\pm0.10)\times 10^{-6}$
and the limit $\mathcal{B}(B^0_s\rightarrow \mu^+\mu^-\phi)/\mathcal{B}(B^0_s\rightarrow J/\psi\phi)<2.61(2.30)\times 10^{-3}$
at 95(90)$\%$ confidence level.
\end{abstract}

\section{Introduction}
In the standard model (SM) both the decay of a $B^0_{s,d}$ meson into two muons and
the decay of a $b$ quark into an $s$ quark and two muons are highly suppressed.
They require a Flavor Changing Neutral Current (FCNC) process which can only
occur through higher order Feynman diagrams.\\
Processes beyond the standard model can contribute to those decays and significantly
modify the decay rates. Any sizable deviation of the measurement from the SM prediction
would indicate new physics phenomena. In these proceedings we present the current searches
of the CDF Collaboration for the branching ratios of the rare decays
$B^0_{(s,d)}\rightarrow\mu^+\mu^-$ \cite{cdf_bsmumu:2007} and $B\rightarrow \mu^+\mu^- h$
\cite{cdf_bmumuh:2006}, where
B stands for $B^+$, $B^0_d$ or $B^0_s$, and h stands for $K^+$, $K^{*0}$ or $\phi$ respectively.

\section{Search for the rare decays $\mathbf{B^0_{(s,d)}\rightarrow\mu^+\mu^-}$}
\begin{figure}
\includegraphics[width=18pc]{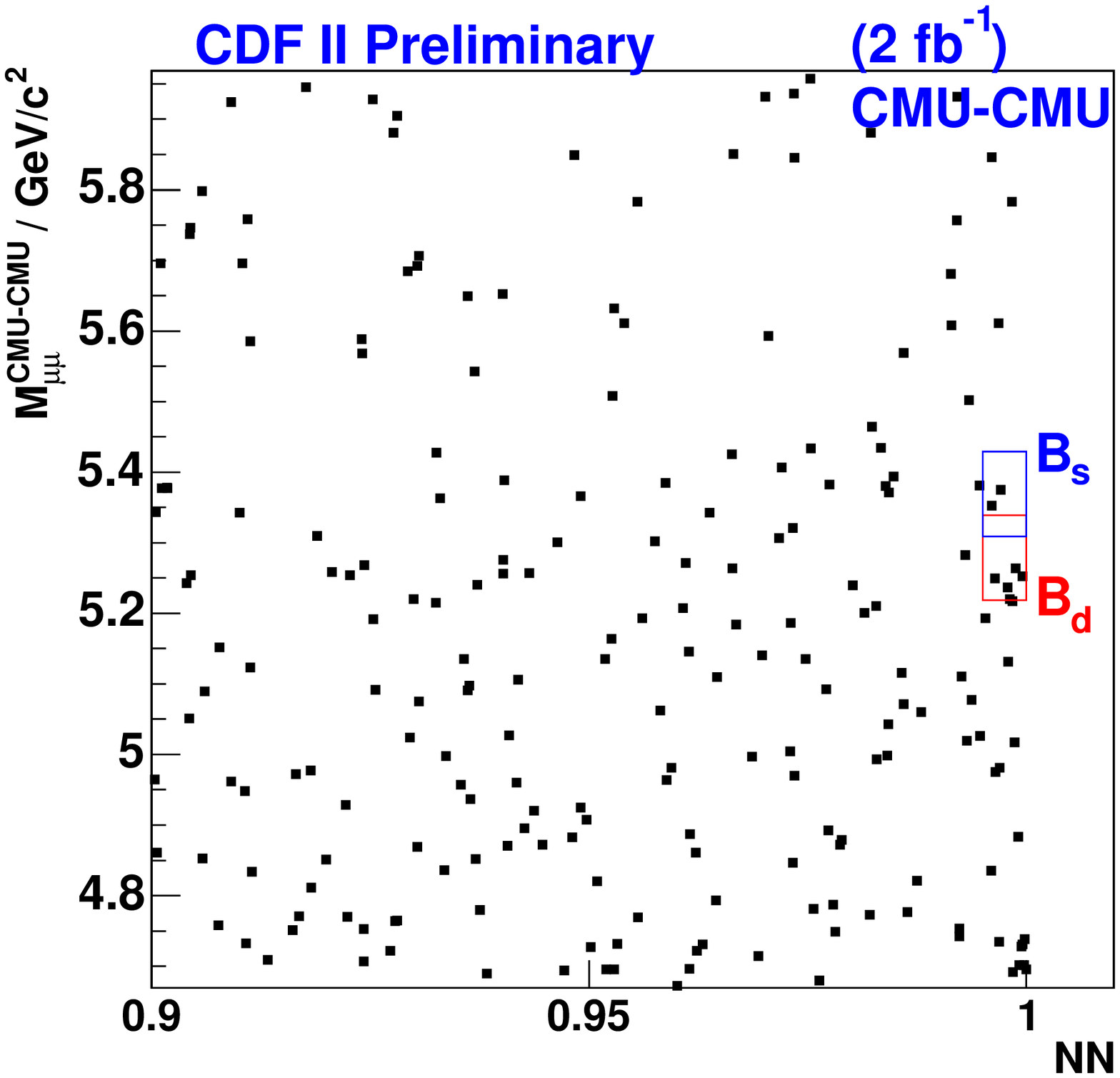}\hspace{2pc}%
\includegraphics[width=18pc]{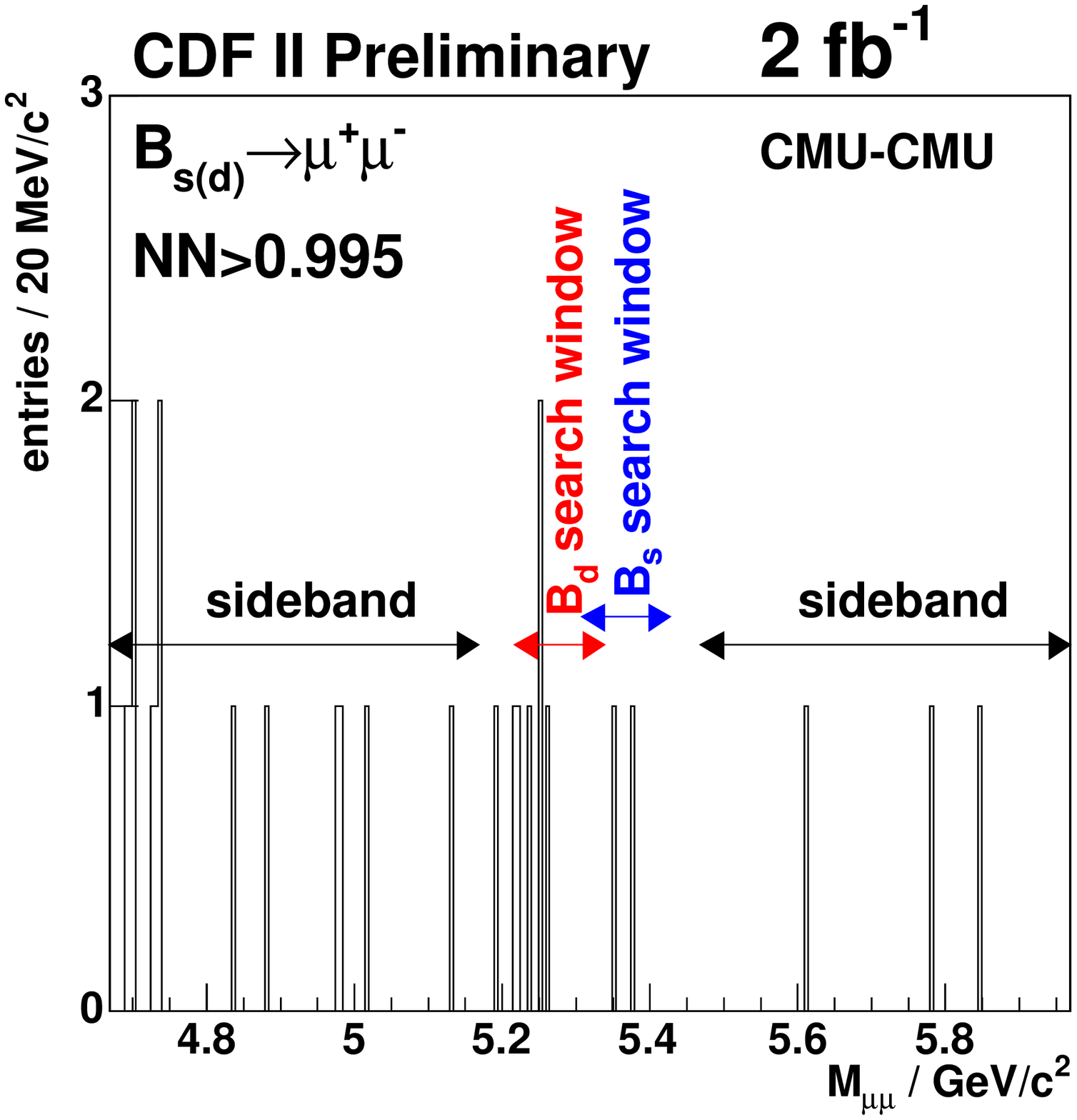}%
\caption{\label{Bs_mumu_selection}The invariant dimuon mass distribution versus the NN output for
events satisfying the baseline cuts in the CMU-CMU mode (left). The NN output bin NN$\ge0.995$ is the
most sensitive search area. Two candidates are found in the $B^0_s$ mass range, five candidates in the
$B^0_d$ mass range. The right plot shows the dimuon mass distribution for events with NN$\ge0.995$.}
\end{figure}

The SM predictions for the branching ratios of the decays 
$B^0_{s,d}\rightarrow\mu^+\mu^-$ are 
of the order of $10^{-10} - 10^{-9}$
\cite{Buras:2003td}.
These low rates can not be probed at the Tevatron yet. However, new physics models
like supersymmetry can enhance those ratios to a level that
can be observed by the Tevatron experiments.\\
The analysis uses 2 fb$^{-1}$ of data \cite{bsmumu_note_eps}, selected by the CDF dimuon trigger.
The CDF muon system consists of several different components, most notably the central muon
system (CMU), covering the pseudorapidity range $|\eta|\le0.6$, and the central muon extension (CMX),
covering $0.6\le|\eta|\le1.0$.
The dimuon trigger system requires at least
one muon in the CMU. The two resulting trigger possibilities, CMU-CMU and CMU-CMX, have different
acceptances and detector efficiencies, so that they will be treated separately during the analysis.\\
After the application of some baseline cuts, the selection of $B^0_{s,d}\rightarrow\mu^+\mu^-$
candidates uses vertex
displacement, transverse momentum and isolation variables that are combined in a neural
network (NN) to form a single variable, discriminating between signal and background.
For network training $B^0_{s,d}\rightarrow\mu^+\mu^-$
signal events from simulation and background events from data sidebands are used.
The invariant dimuon mass distribution is shown in Fig. \ref{Bs_mumu_selection}.

The optimal search strategy is determined using the \emph{a priori} expected 95\% C.L. upper limit
as figure of merit. The search window is divided into 15 regions -- 5 mass bins times 3 NN output
bins.
The observed number of events is compatible with the estimated background.
Combinatoric background is estimated by sideband interpolation and
the contribution from misidentified hadronic $B^0_{s,d}$ decays is obtained
by a combination of data and simulation.
The limit on the branching fraction is obtained by normalizing to the well-measured
$B^+\rightarrow J/\psi K^+$ decay. After combining all 15 regions, the CMU-CMU and CMU-CMX modes
and using the world average value for $\mathcal{B}(B^+\rightarrow J/\psi K^+)$ \cite{Yao:2006px}, we obtain
the $90\%(95\%)$ C.L. limits of
$\mathcal{B}(B^0_s\rightarrow \mu^+\mu^-)<4.7\times 10^{-8} (5.8\times 10^{-8})$ and
$\mathcal{B}(B^0_d\rightarrow \mu^+\mu^-)<1.5\times 10^{-8} (1.8\times 10^{-8})$. These results are the
most stringent limits to date.

\section{Search for the rare decays $\mathbf{B^+\rightarrow\mu^+\mu^-K^+}$,
 $\mathbf{B^0_d\rightarrow\mu^+\mu^-K^{\ast 0}}$ and $\mathbf{B^0_s\rightarrow\mu^+\mu^-\phi}$ }
The transition $b \rightarrow s \mu^+\mu^-$ is sensitive to new physics since
physics beyond the SM can manifest itself in significant deviations from
SM transition rate predictions.
The analysis uses a data sample corresponding to 924 pb$^{-1}$. Each branching fraction
is determined by the corresponding resonant $B\rightarrow J/\psi h$ mode for normalization,
resulting in a cancellation of many systematic uncertainties. After some baseline cuts and
the removal of resonant dimuon contributions
the selection is optimized using the following variables: the proper decay time significance $ct/\sigma_{ct}$,
the pointing angle from the $B$ meson candidate to the primary vertex and the isolation of the
$B$ meson candidate. The figure of merit in the optimization process is $S/\sqrt{S+B}$, with
$S$ the expected signal yield, determined from the normalization modes, and $B$ the expected
background yield. The following background sources are considered: misidentification of hadrons as muons,
reflections between the three modes, and combinatorial background.

\begin{figure}
\includegraphics[width=12pc]{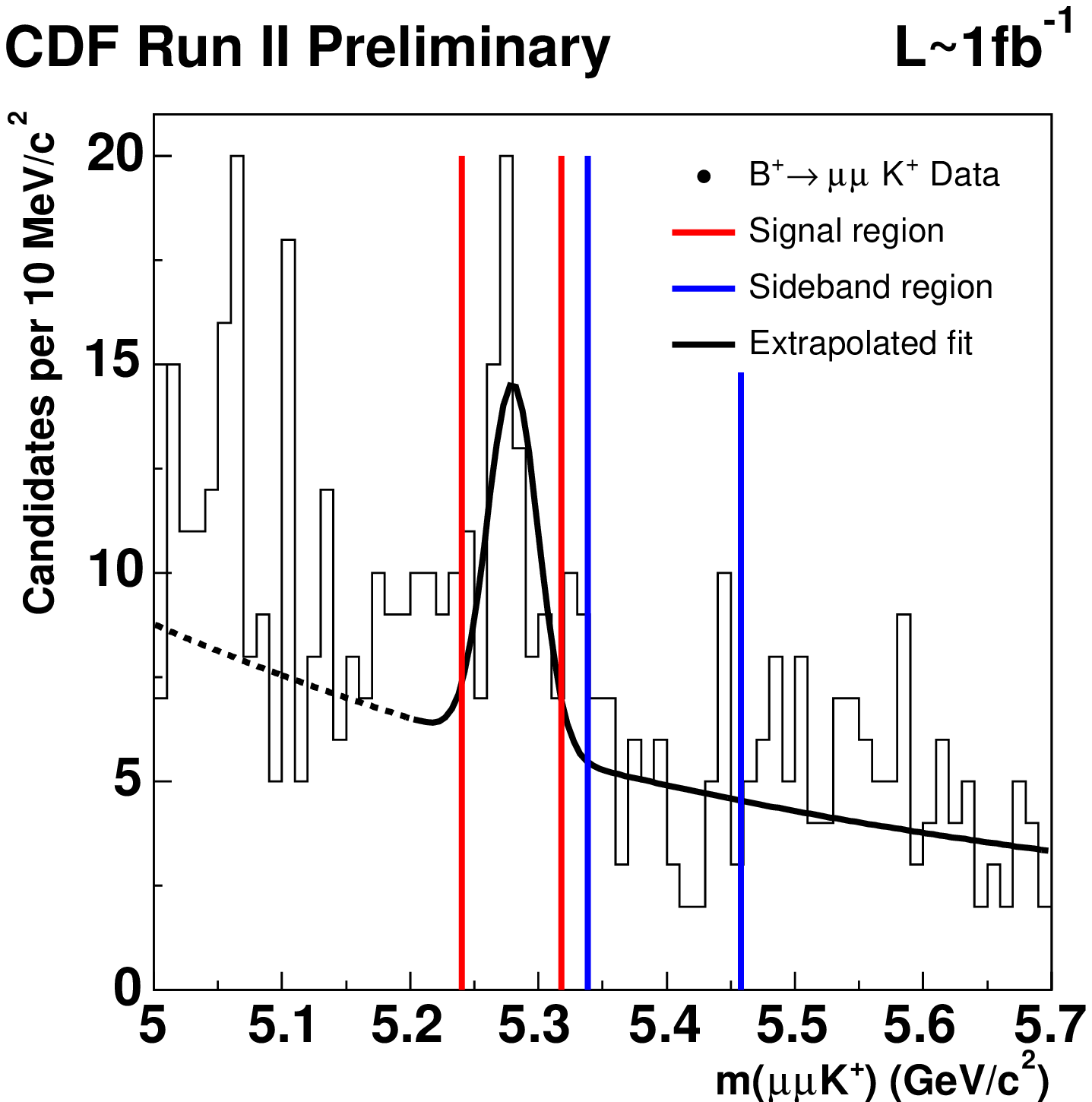}\hspace{1pc}%
\includegraphics[width=12pc]{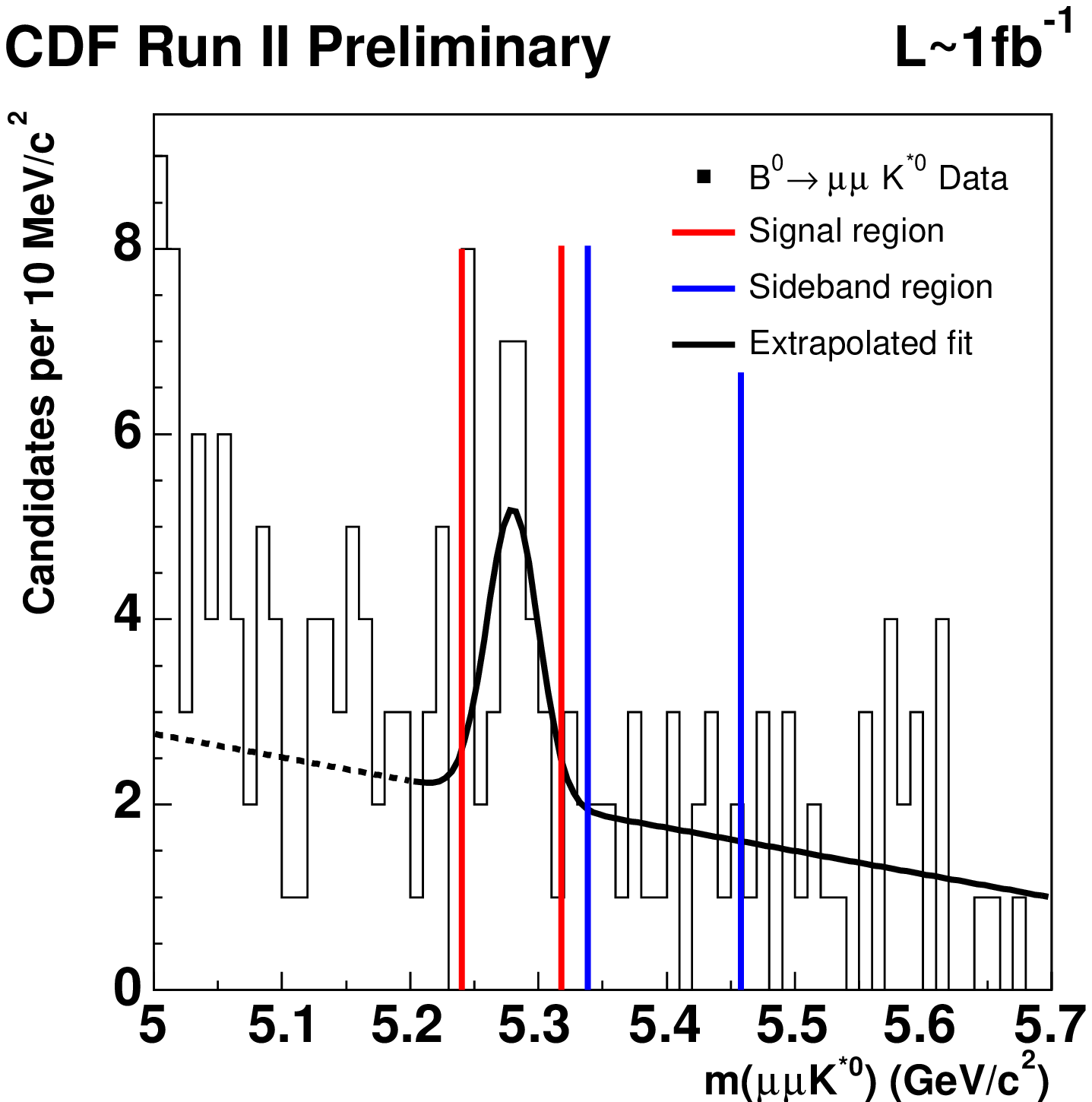}\hspace{1pc}%
\includegraphics[width=12pc]{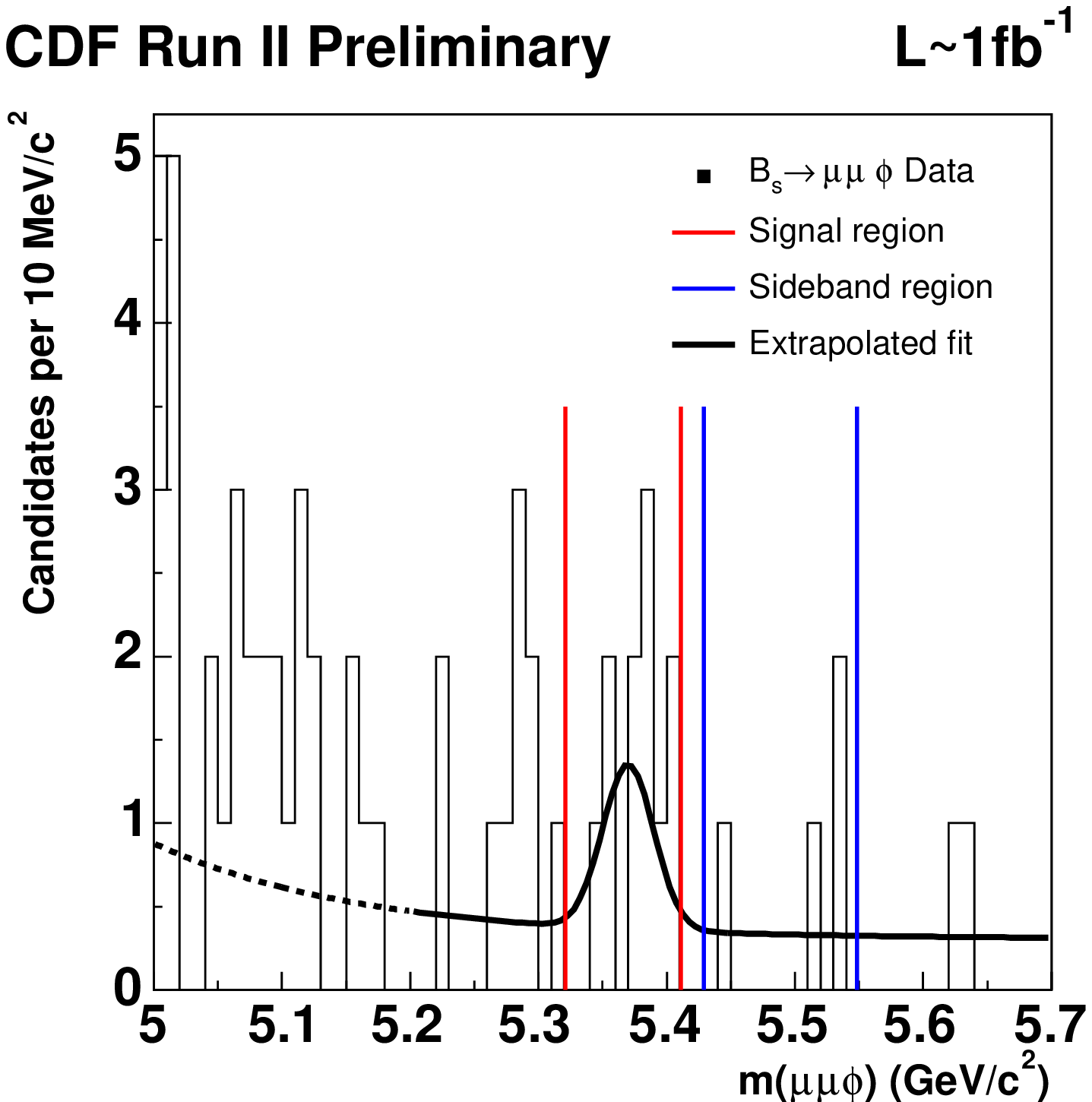}%
\caption{\label{B_mumuh_selection}Invariant mass spectra for the rare modes. The black curve illustrates
the expected shape after background estimation and determination of the signal yield. The dotted part of
the curve indicates that it is not expected to describe the low mass background, containing partially
reconstructed B decays.}
\end{figure}
The three final invariant mass distributions are shown in Fig. \ref{B_mumuh_selection}. Each
mode shows an excess in the signal mass region. The significance of the excess is determined by
calculating the probability for the expected background to fluctuate into at least the number of observed
events.  From the event counts, together with the normalization sample,
the relative branching fractions between signal mode and normalization mode are obtained. The absolute
branching fractions are calculated with the world average branching fraction values for the normalization
modes \cite{Yao:2006px}. Table \ref{mumuh_results} summarizes the results.

\begin{table}[h]
\caption{\label{mumuh_results}Summary of the $B\rightarrow \mu^+\mu^- h$ results. Listed for each decay mode are the
number of observed signal events, the number of expected background events and the signal significance.}
\begin{center}
\lineup
\begin{tabular}{lccc}
\br
Mode                       & $B^+\rightarrow\mu^+\mu^-K^+$ & $B^0_d\rightarrow\mu^+\mu^-K^{\ast 0}$ & $B^0_s\rightarrow\mu^+\mu^-\phi$ \\
\mr
N observed events          &                          90  &                                    35  &                            11    \\
N expected BG events       &           $45.3 \pm 5.8$     &                     $16.5 \pm 3.6$     &               $3.5 \pm 1.5$      \\
Gaussian significance      &               4.5            &                     2.9                &                 2.4              \\
\mr
Rel $\mathcal{B} \pm stat \pm syst \times 10^{-3}$
                           &        $0.59\pm0.15\pm0.03$  &                  $0.62\pm0.23\pm0.07$  &          $1.24\pm0.60\pm0.15$    \\
Abs $\mathcal{B} \pm stat \pm syst \times 10^{-6}$
                           &        $0.60\pm0.15\pm0.04$  &                  $0.82\pm0.31\pm0.10$  &          $1.16\pm0.56\pm0.42$    \\
Rel $\mathcal{B}$ 90\%CL limit $\times 10^{-3}$
                           &           -                  &                  -                     &           2.30                   \\
Rel $\mathcal{B}$ 95\%CL limit $\times 10^{-3}$
                           &         -                    &                  -                     &           2.61                   \\
\br
\end{tabular}
\end{center}
\end{table}

The results for
$\mathcal{B}(B^+\rightarrow\mu^+\mu^-K^+)$ and 
$\mathcal{B}(B^0_d\rightarrow\mu^+\mu^-K^{\ast 0})$ are in good agreement with previous measurements
\cite{Ishikawa:2006fh,Aubert:2006vb} and SM predictions \cite{Ali:1999mm}. The limit
on $\mathcal{B}(B^0_s\rightarrow\mu^+\mu^-\phi)$ is the most stringent to date.

\section*{References}
\bibliography{EPS_rareDecays_CDF_bib}

\end{document}